\documentstyle[graphics,aps]{revtex}

\def\mat#1#2#3{\langle{#1}\vert{#2}\vert{#3}\rangle}

\begin{document}

\title{
\hfill
\parbox{3cm}{\normalsize DPNU-01-13}\\
\vspace{0.5cm}
{\bf $B\to\phi K$ decays in perturbative QCD approach} }
\author
{S.~Mishima\footnote{e-mail: mishima@eken.phys.nagoya-u.ac.jp}\\
{\normalsize \it Department of Physics, Nagoya University, Nagoya
464-8602, Japan}
}
%\date{}

\maketitle
\vspace{10mm}

\begin{abstract}
We calculate the branching ratios and CP asymmetries of the $B\to \phi K$ 
decays using perturbative QCD approach, which includes $k_T$ and
 threshold resummations. Our results of branching ratios are consistent
 with the experimental data and larger than those obtained from the
 naive factorization assumption and the QCD-improved factorization approach.   
\end{abstract}

\vspace{10mm}

%%%%%%%%%%%%%%%%%%%%%%%%%%%%%%%%%%%%%%%%%%%%%%%%%%%%%%
%
%            Introduction
%
%%%%%%%%%%%%%%%%%%%%%%%%%%%%%%%%%%%%%%%%%%%%%%%%%%%%%%

\section{Introduction}
Recently the branching ratios of the $B\to \phi K$ decays have been
measured by the BaBar\cite{BABAR}, BELLE\cite{BELLE} and CLEO\cite{CLEO}
collaborations.  
These decay modes are
important to understand penguin dynamics, because the $B\to \phi K$
modes are pure penguin processes. 

There is a following problem related a penguin contribution for
amplitude\cite{PEN}.  
A naive estimate of the loop diagram leads to $P/T \sim \alpha_s/(12\pi)
\log(m_t^2/m_c^2)\sim O(0.01)$ where $P$ is a penguin amplitude and $T$
is a tree amplitude. But experimental data for ${\rm Br}(B \to K\pi)$ and
${\rm Br}(B \to \pi\pi)$ leads to $P/T\sim O(0.1)$. 
The penguin contribution is enhanced dynamically.
This enhancement can not be explained by factorization
assumption for nonleptonic two-body $B$ meson decays\cite{BSW}. 
This problem is explained by Keum, Li and Sanda using
perturbative QCD (PQCD) approach\cite{KLS}. 

PQCD method for inclusive decays
was developed by many researcher over many years, and
this formalism has been successful. Recently, PQCD has been applied to
the exclusive B meson decays,
$B \to K\pi$\cite{KLS}, 
$\pi\pi$\cite{LUY}, $\pi\rho$, $\pi\omega$\cite{LY},
$KK$\cite{KK} and $K\eta^{(')}$\cite{KoSa}.
We calculate the branching ratios for
the $B\to \phi K$ modes using this approach to leading order in
$\alpha_s$ and leading power in $1/M_B$ where $M_B$ is the $B$ meson
mass. 
The $B\to \phi K$ modes depend only on penguin amplitudes, 
and the branching ratio for this modes is proportional to $|P|^2$. 
Thus the $B\to \phi K$ modes are useful to seeing the size of the
penguin amplitudes.  

In this paper we compute the branching ratio for the $B\to \phi K$
modes using PQCD approach. In Sec.~\ref{sec:PF}, we introduce the
PQCD formalism for exclusive B meson decays. In
Sec.~\ref{sec:amp}, we calculate analytic formula
of the factorizable amplitudes and the nonfactorizable amplitudes of
the decays. In Sec.~\ref{sec:num}, we predict the branching
ratios of the decays and the direct CP asymmetry of the charged
mode. Our predicted branching ratios agree with the current experimental
data and are larger than the values obtained from the naive factorization
assumption and the QCD-improved factorization\cite{QIF1}\cite{QIF2}.
In the Appendix, we review the meson wave functions with up to twist-3
terms which are proportional to $m_{0K}/M_B$ or $M_\phi/M_B$.

%%%%%%%%%%%%%%%%%%%%%%%%%%%%%%%%%%%%%%%%%%%%%%%%%%%%%%
%
%            PQCD
%
%%%%%%%%%%%%%%%%%%%%%%%%%%%%%%%%%%%%%%%%%%%%%%%%%%%%%%

\section{PQCD Factorization Theorem}\label{sec:PF}
The effective Hamiltonian for $\Delta S = 1$ transitions is
\begin{equation}
H_{\rm eff}={G_F\over\sqrt{2}}
\sum_{q=u,c}V_{qs}^* V_{qb}\left[C_1(\mu)O_1^{(q)}(\mu)
+C_2(\mu)O_2^{(q)}(\mu)+\sum_{i=3}^{10}C_i(\mu)O_i(\mu)\right]\;, 
\label{hbk}
\end{equation}
where $V_{qs}^*$ and $V_{qb}$ are the Cabibbo-Kobayashi-Maskawa matrix
elements\cite{KoMa} 
and $O_{1-10}$ are local four-fermi operators.
The local operators are given by
\begin{eqnarray}
& &O_1^{(q)} = (\bar{s}_iq_j)_{V-A}(\bar{q}_jb_i)_{V-A}\;,
\;\;\;\;\;\;\;\;\;\;\;
O_2^{(q)} = (\bar{s}_iq_i)_{V-A}(\bar{q}_jb_j)_{V-A}\;, 
\nonumber \\
& &O_3 =(\bar{s}_ib_i)_{V-A}\sum_{q}(\bar{q}_jq_j)_{V-A}\;,
\;\;\;\;\;\;\;\;
O_4 =(\bar{s}_ib_j)_{V-A}\sum_{q}(\bar{q}_jq_i)_{V-A}\;, 
\nonumber \\
& &O_5 =(\bar{s}_ib_i)_{V-A}\sum_{q}(\bar{q}_jq_j)_{V+A}\;,
\;\;\;\;\;\;\;\;
O_6 =(\bar{s}_ib_j)_{V-A}\sum_{q}(\bar{q}_jq_i)_{V+A}\;, 
\nonumber \\
& &O_7 =\frac{3}{2}(\bar{s}_ib_i)_{V-A}\sum_{q}e_q(\bar{q}_jq_j)_{V+A}\;,
\;\;
O_8 =\frac{3}{2}(\bar{s}_ib_j)_{V-A}\sum_{q}e_q(\bar{q}_jq_i)_{V+A}\;, 
\nonumber \\
& &O_9 =\frac{3}{2}(\bar{s}_ib_i)_{V-A}\sum_{q}e_q(\bar{q}_jq_j)_{V-A}\;,
\;\;
O_{10} =\frac{3}{2}(\bar{s}_ib_j)_{V-A}\sum_{q}e_q(\bar{q}_jq_i)_{V-A}\;, 
\end{eqnarray} 
where $i, \ j$ is the color indices and $q$ is taken $u,\;d,\;s$ and $c$.
The Wilson coefficients $C_{1-10}$ are calculated up to next-to-leading
order by Buras {\it et al}.\cite{REVIEW}. 

\begin{figure}[hbt]
\begin{center}
\includegraphics{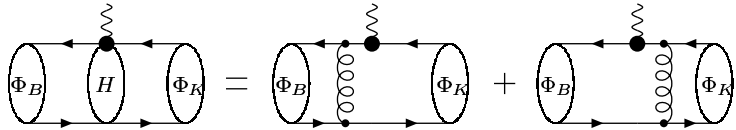}
\caption{The $B \to K$ transition form factor 
in leading order of $\alpha_s$}
\label{fig:lowest}
\end{center}
\end{figure}

PQCD approach is
based on the three scale factorization theorem\cite{CL}\cite{YL} in
which the matrix 
element can be written as the convolution of the hard part $H$ with the
hard scale $t\sim O(M_B)$, the meson wave
function $\Phi$ with the soft scale $1/b\sim \Lambda_{\rm QCD}$ and the 
Wilson coefficient $C$ whose evolves from the $W$ boson mass $M_W$
down to the scale $t$. 
$b$ is the conjugate variable of the parton transverse momenta $k_T$. 
For instance, the $B \to K$ transition form factor is written as
\begin{equation}
F^{BK} \sim \int [dx][db] C(t) \Phi_K(x_2,b_2) H(t)\Phi_B(x_1,b_1)
\exp\left[-\sum_{i=1,2}\int_{1/b_i}^t\frac{d{\bar\mu}}
{\bar\mu}\gamma_\phi(\alpha_s({\bar\mu}))\right]\;,
\end{equation}
where $x$ is the momentum fraction of parton and $\gamma_\phi$ is the
anomalous dimension of meson.
The hard part $H$ can be calculated perturbatively and 
the Feynman diagrams are described as Fig.~\ref{fig:lowest} 
in leading order of $\alpha_s$. 
%%%%%%%%%%%%%%% Fig. 1  %%%%%%%%%%%%%%%%%%%%%%%%%%%%%%%%%%%%%%%
%%%%%%%%%%%%%%%%%%%%%%%%%%%%%%%%%%%%%%%%%%%%%%%%%%%%%%%%%%%%%%%
It is important that the scale of the Wilson coefficient is determined
dynamically, since the scale depends on $x$ and $b$.   
Then in PQCD, the scale of the Wilson coefficient can reach lower scale
than $M_B/2$ which are usually taken in the naive factorization
assumption case.  
The Wilson coefficients for the penguin operator are enhanced as
the scale evolves to lower scale.
Therefore, the penguin amplitudes are larger than those in the naive
factorization assumption. 

In this method, we can calculate not only factorizable amplitudes but
also nonfactorizable and annihilation amplitudes which can not be
calculated in the factorization assumption. 

\begin{figure}[hbt]
\begin{center}
\includegraphics{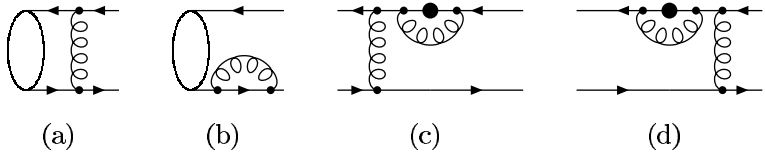}
\end{center}
\caption{These are the diagrams which generate double logarithms
 corrections 
 in infrared divergences. Fig.~\ref{fig:resum}(a) and
 Fig.~\ref{fig:resum}(b) are the correction of the meson
 wave function. Fig.~\ref{fig:resum}(c) and Fig.~\ref{fig:resum}(c) 
are the correction of the hard part.} 
\label{fig:resum}
\end{figure}

This $B \to K$ transition form factor has two types of double
logarithmic corrections in infrared divergences.
The one is $\alpha_s\ln^2(Pb)$ from the overlap of
collinear and soft divergences which are generated by the correction of
the meson wave function as Fig.~\ref{fig:resum}(a) and
Fig.~\ref{fig:resum}(b). The resummation of this double logarithms 
leads to the factor $\exp[-s(P,b)]$\cite{CS}\cite{BS}. The explicit
form of this factor is shown, for example, in Ref.~\cite{L1}.
This resummation is called $k_T$ resummation. 
The other double logarithm is $\alpha_s\ln^2(1/x)$ from the end-point
region of the momentum fraction $x$\cite{TR}. This double logarithm is
generated by the corrections of the hard part as 
Fig.~\ref{fig:resum}(c) and Fig.~\ref{fig:resum}(d). 
The resummation of this double logarithms is
called the threshold resummation.   
We use the approximate form of this resummation factor that is $T(x)=N_t
\{x(1-x)\}^c$ where $N_t=1.775$ and $c=0.3$\cite{KuLS}. 
The Sudakov factor from $k_T$ and threshold resummations suppress large
$b$ region and end-point of $x$ where the hard part $H$ is singular, the
 hard part $H$ can then be calculated perturbatively.
%%%%%%%%%%%%%%% Fig. 2 %%%%%%%%%%%%%%%%%%%%%%%%%%%%%%%%%%%%%%%
%%%%%%%%%%%%%%%%%%%%%%%%%%%%%%%%%%%%%%%%%%%%%%%%%%%%%%%%%%%%%%

In PQCD approach, we can calculate hard part
perturbatively, and we use the meson wave functions which are formed in
the light-cone QCD sum rules. 
The hard part depends on processes, but the wave
functions are independent of those. We can determine the parameters
in the wave functions, since there are many modes in the $B$ meson
 two-body decays. So, we can predict quantitative values for the other
 decays.  
 
In this study, we calculate the hard part in leading order of
$\alpha_s$. The meson wave functions are used with
leading twist and twist-3 terms which are proportional to $m_{0K}/M_B$
or $M_\phi/M_B$.  
We review the meson wave functions in the Appendix.

%%%%%%%%%%%%%%%%%%%%%%%%%%%%%%%%%%%%%%%%%%%%%%%%%%%%%%%%%%%
%
%        $B\to \phi K$ Amplitudes
%
%%%%%%%%%%%%%%%%%%%%%%%%%%%%%%%%%%%%%%%%%%%%%%%%%%%%%%%%%%%

\section{$B\to \phi K$ Amplitudes}\label{sec:amp}
\begin{figure}[hbt]
\begin{center}
\includegraphics{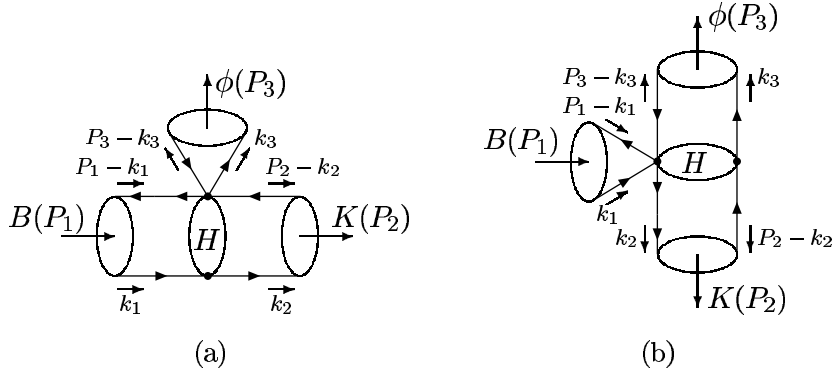}
\end{center}
\caption{Momentum assignment}
\label{fig:mom}
\end{figure}

In the light-cone coordinate, the $B$ meson momentum $P_1$, the $K$
meson momentum $P_2$ and the $\phi$ meson momentum $P_3$ are taken to be
\begin{eqnarray}
P_1=\frac{M_B}{{\sqrt 2}}(1,1,{\bf 0}_T)\;,\;\;
P_2=\frac{M_B}{{\sqrt 2}}(1-r_{\phi}^2,0,{\bf 0}_T)\;,\;\;
P_3=\frac{M_B}{{\sqrt 2}}(r_{\phi}^2,1,{\bf 0}_T)\;,
\end{eqnarray}
where $r_{\phi} = M_{\phi}/M_B$, and the $K$ meson mass is neglected. 
In this situation, the $B$ meson is at rest.
The quark and anti-quark momenta
in the mesons are displayed in Fig.~\ref{fig:mom}. Their components are
\begin{eqnarray}
k_1&=&(0,x_1P_1^-,{\bf k}_{1T}),\;\;\;\;\;
P_1-k_1=(P_1^+,(1-x_1)P_1^-,-{\bf k}_{1T})\;, \nonumber\\
k_2&=&(x_2P_2^+,0,{\bf k}_{2T}),\;\;\;\;\;
P_2-k_2=((1-x_2)P_2^+,0,-{\bf k}_{2T})\;,  \nonumber\\
k_3&=&(0,x_3P_3^-,{\bf k}_{3T}),\;\;\;\;\;
P_3-k_3=(0,(1-x_3)P_3^-,-{\bf k}_{3T})\;.
\end{eqnarray}
We can integrate over $k_1^+$ and choose $k_1^+=0$,  
since the hard part is independent of $k_1^+$.
%%%%%%%%%%%%%%% Fig. 3 %%%%%%%%%%%%%%%%%%%%%%%%%%%%%%%%%%%%%%%
%%%%%%%%%%%%%%%%%%%%%%%%%%%%%%%%%%%%%%%%%%%%%%%%%%%%%%%%%%%%%%
The $\phi$ meson longitudinal polarization vector $\epsilon_{\phi}$ and
two transverse polarization vector $\epsilon_{\phi T}$ are given by 
\begin{eqnarray}
\epsilon_{\phi}=\frac{1}{\sqrt{2}r_{\phi}}( -r_{\phi}^2, 1, {\bf 0}_T)
\;,\;\;\;\;\; 
\epsilon_{\phi T}=(0,0,{\bf 1}_T)\;.
\end{eqnarray}

The $B$ meson wave function for incoming state and the $K$ and $\phi$
meson wave functions for outgoing state with up to twist-3 terms are
written as  
\begin{eqnarray}
\Phi^{(in)}_{B,\alpha\beta,ij}&=&
\frac{i \delta_{ij}}{\sqrt{2N_c}}\int dx_1 d^2{{\bf k}_{1T}} 
e^{-i(x_1P_1^-z_1^+-{{\bf k}_{1T}}{{\bf z_1}_T})}
\left[
(\not P_1 + M_B)\gamma_5 \phi_B(x_1,{\bf k}_{1T})
\right]_{\alpha\beta}\;,\\
\Phi^{(out)}_{K,\alpha\beta,ij}
&=&\frac{-i\delta_{ij}}{\sqrt{2N_c}}\int^1_0dx_2 e^{ix_2P_2\cdot z_2}
\gamma_5\left[
\not P_2\phi^A_K(x_2)+m_{0K}\phi_K^P(x_2)
+m_{0K} (\not v\not n -1)\phi_K^T(x_2)
\right]_{\alpha\beta}\;,\\
\Phi^{(out)}_{\phi,\alpha\beta,ij}
&=&\frac{\delta_{ij}}{\sqrt{2N_c}}\int^1_0dx_3 e^{ix_3P_3\cdot z_3}
\left[
M_\phi\not\epsilon_\phi \phi_\phi(x_3)+
\not\epsilon_\phi \not P_3\phi_\phi^t(x_3)
+ M_\phi\phi_\phi^s(x_3)
\right]_{\alpha\beta}\;,
\end{eqnarray}
where $i$ and $j$ is the color indices, $\alpha$ and $\beta$ are the
Dirac indices. $m_{0K}$ is related to chiral symmetry breaking scale, 
$m_{0K}= M_K^2/(m_d+m_s)$. $v$ and $n$ are defined as $v^\mu =
P_2^\mu/P_2^+$ and $n^\mu = z_2^\mu/z_2^-=(0,1,{\bf 0}_T)$.
We neglect the wave functions which are proportional to the transverse
polarization vector $\epsilon_\phi^T$, because these terms do not appear
in our calculations. 
It must be noted that the expression of $\Phi^{(out)}_{\phi}$ 
depends on how to define the 
sign of $\epsilon_\phi$.
The explicit form of these wave functions will be shown in
Sec. \ref{sec:num} and the detail is shown in Appendix.

The decay rates for the $B\to \phi K$ have the expressions
\begin{equation}
\Gamma=\frac{G_F^2}{32\pi M_B}|{\cal A}|^2\;.
\end{equation}
The decay amplitudes ${\cal A}$ and $\bar{\cal A}$ corresponding to
$B^0\to \phi K^0$ and ${\bar B^0}\to \phi {\bar K^0}$ respectively, are
written as 
\begin{eqnarray}
{\cal A}&=&f_{\phi}V_{ts}V^*_{tb}
F^P_e+V_{ts}V^*_{tb}
{\cal M}^P_e
+f_BV_{ts}V^*_{tb}
F^P_a+V_{ts}V^*_{tb}
{\cal M}^P_a\;,\\
\bar{\cal A}&=&f_{\phi}V^*_{ts}V_{tb}F^P_e+V^*_{ts}V_{tb}{\cal M}^P_e
+f_BV^*_{ts}V_{tb}F^P_a+V^*_{ts}V_{tb}{\cal M}^P_a\;.
\end{eqnarray}
$F_e$ is the amplitude for the factorizable diagrams which are
considered in the factorization assumption.
$F_a$ and ${\cal M}$ are the annihilation factorizable and the
nonfactorizable diagrams which are neglected in the factorization
assumption. 
The indices $e$ and $a$ denote the tree topology as
Fig.~\ref{fig:mom}(a) and annihilation topology as 
Fig.~\ref{fig:mom}(b) respectively.
The index $P$ denotes the contribution from the diagram with a penguin
operator.   
The decay amplitudes ${\cal A}^+$ and ${\cal A}^-$ corresponding to
$B^+\to \phi K^+$ and $B^-\to \phi K^-$ respectively, are written as
\begin{eqnarray}
{\cal A}^+&=&f_{\phi}V_{ts}V^*_{tb}
F^P_e+V_{ts}V^*_{tb}
{\cal M}^P_e
+f_BV_{ts}V^*_{tb}
F^P_a+V_{ts}V^*_{tb}
{\cal M}^P_a
-f_BV_{us}V^*_{ub}
F^T_a-V_{us}V^*_{ub}
{\cal M}^T_a\;,\\
{\cal A}^-&=&f_{\phi}V^*_{ts}V_{tb}F^P_e+V^*_{ts}V_{tb}{\cal M}^P_e
+f_BV^*_{ts}V_{tb}F^P_a+V^*_{ts}V_{tb}{\cal M}^P_a
-f_BV^*_{us}V_{ub}F_a^T-V^*_{us}V_{ub}{\cal M}^T_a\;,
\end{eqnarray}
where there are very small contribution from tree diagrams, 
the index $T$ denotes these contribution.

\subsection{Factorizable Amplitudes}\label{sec:fa}
%%%%%%%%%%%%%%% Fig. 4 %%%%%%%%%%%%%%%%%%%%%%%%%%%%%%%%%%%%%%%
\begin{figure}[hbt]
\begin{center}
\includegraphics{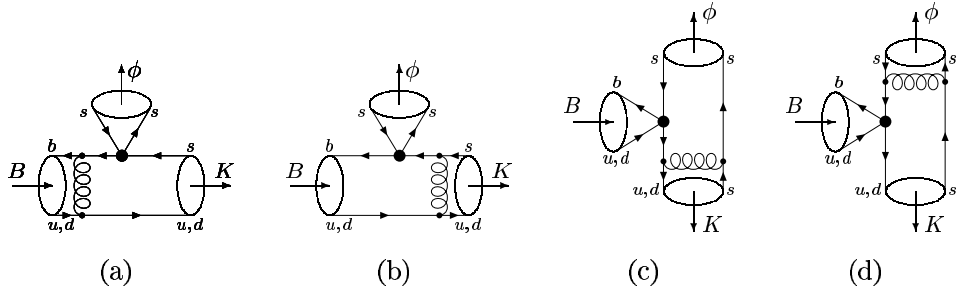}
\end{center}
\caption{Feynman diagrams contributing to factorizable amplitudes}
\label{fig:fact}
\end{figure}
%%%%%%%%%%%%%%%%%%%%%%%%%%%%%%%%%%%%%%%%%%%%%%%%%%%%%%%%%%%%%%
The factorizable amplitudes which come from
Fig.~\ref{fig:fact} are written as
\begin{eqnarray}
F^P_e &=& F^P_{3e}+F^P_{4e}+F^P_{5e} \nonumber\\ 
&=&
8\pi C_F M_B^4 \int_0^1 dx_1 dx_2 \int_0^{\infty} b_1db_1 b_2db_2 
\phi_B(x_1,b_1)
\nonumber \\
& &\times
\bigg\{ 
\left[ (1+x_2)\phi_K^A(x_2)+r_K(1-2x_2)
\left( \phi_K^P(x_2)+\phi_K^T(x_2) \right) \right]
\nonumber\\
& & \;\;\;\;\;\;\times
E_{e}(t^{(1)}_e) N_t \{x_2(1-x_2)\}^c h_e(x_1,x_2,b_1,b_2)
\nonumber\\
& &\;\;\;\;\;\; +2r_K \phi_K^P(x_2)E_{e}(t^{(2)}_e) 
N_t \{x_1(1-x_1)\}^c
h_e(x_2,x_1,b_2,b_1)
\bigg\}\;,\\
F^P_a&=&F^P_{a4}+F^P_{a6}\;,
\nonumber\\
F^P_{a4}&=&
8\pi C_F M_B^4 \int_0^1 dx_2 dx_3 \int_0^{\infty} b_2db_2 b_3db_3 
\nonumber \\
& &\times
\bigg\{ 
\left[ 
x_3\phi_K^A(x_2)\phi_{\phi}(x_3)
-2r_Kr_\phi \phi_K^P(x_2)
\left(
\left( \phi_{\phi}^t(x_3) - \phi_{\phi}^s(x_3) \right)
- x_3\left( \phi_{\phi}^t(x_3) + \phi_{\phi}^s(x_3) \right)
\right)
\right]
\nonumber \\
& &\;\;\;\;\;\;\times
E_{a4}(t^{(1)}_a)  N_t \{x_3(1-x_3)\}^c h_a(1-x_2,x_3,b_2,b_3)
\nonumber\\
& &\;\;\;\;\;\; 
 - \left[
(1-x_2)\phi_K^A(x_2)\phi_{\phi}(x_3)
+2r_Kr_\phi
\left(
2\phi_K^P(x_2)
- x_2\left( \phi_K^P(x_2)-\phi_K^T(x_2) \right)
\right) \phi_{\phi}^s(x_3)
\right]
\nonumber \\
& &\;\;\;\;\;\;\times
E_{a4}(t^{(2)}_a) N_t \{x_2(1-x_2)\}^c h_a(x_3,1-x_2,b_3,b_2)
\bigg\}\;,\\
F^P_{a6}&=&
16\pi C_F M_B^4 \int_0^1 dx_2 dx_3 \int_0^{\infty} b_2db_2 b_3db_3 
\nonumber \\
& &\times
\bigg\{ 
\left[ 2 r_K \phi_K^P(x_2)\phi_{\phi}(x_3) 
- r_{\phi}x_3 \phi_K^A(x_2) 
\left( \phi_{\phi}^t(x_3) - \phi_{\phi}^s(x_3) \right) \right]
\nonumber\\
& & \;\;\;\;\;\;\times
E_{a6}(t^{(1)}_a)  N_t \{x_3(1-x_3)\}^c h_a(1-x_2,x_3,b_2,b_3)
\nonumber\\
& &\;\;\;\;\;\;  + \left[ 
r_K (1-x_2)\left( \phi_K^P(x_2)+\phi_K^T(x_2) \right)\phi_{\phi}(x_3)
+ 2 r_{\phi} \phi_K^A(x_2) \phi_{\phi}^s(x_3) \right]
\nonumber\\
& & \;\;\;\;\;\;\times
E_{a6}(t^{(2)}_a) N_t \{x_2(1-x_2)\}^c h_a(x_3,1-x_2,b_3,b_2)
\bigg\}\;.
\end{eqnarray}
The tree amplitude $F_a^T$ is the same as $F_{a4}^P$, but the Wilson
coefficient is different.  
$N_t \{x(1-x)\}^c$ is the factor for the threshold resummation. 
$N_t$ is a normalization factor and $c$ is a constant, we use
$N_t=1.775$ and $c=0.3$.
The evolution factors are defined by 
\begin{eqnarray}
E_{e}(t)=\alpha_s(t)a_e(t)\exp[-S_B(t)-S_K(t)]\;,\;\;
E_{ai}(t)=\alpha_s(t)a_{ai}(t)\exp[-S_K(t)-S_{\phi}(t)]\;.
\end{eqnarray}
The explicit forms of the factor $S_i(t)$ are given, for example, in
Ref.~\cite{KLS}.   
The hard scales $t$, which are the typical scales in hard process, are
given by  
\begin{eqnarray}
t^{(1)}_e&=&{\rm max}(\sqrt{x_2}M_B,1/b_1,1/b_2)\;,
\;\;
t^{(2)}_e={\rm max}(\sqrt{x_1}M_B,1/b_1,1/b_2)\;,
\nonumber\\
t^{(1)}_a&=&{\rm max}(\sqrt{x_3}M_B,1/b_2,1/b_3)\;,
\;\;
t^{(2)}_a={\rm max}(\sqrt{1-x_2}M_B,1/b_2,1/b_3)\;.
\end{eqnarray}
The Wilson coefficients are given by
\begin{eqnarray}
a_e(t)&=& 
\left( C_3+\frac{C_4}{N_c} \right)
+\left( C_4+\frac{C_3}{N_c} \right)
+\left( C_5+\frac{C_6}{N_c} \right) \nonumber\\
& &-\frac{1}{2}\left( C_7+\frac{C_8}{N_c} \right)
-\frac{1}{2}\left( C_9+\frac{C_{10}}{N_c} \right)
-\frac{1}{2}\left( C_{10}+\frac{C_9}{N_c} \right)\;,
\nonumber\\
a_{a4}(t)&=&
\left( C_4+\frac{C_3}{N_c} \right)
+\frac{3}{2}e_q\left( C_{10}+\frac{C_9}{N_c} \right) \;,
\nonumber\\
a_{a6}(t)&=&
\left( C_6+\frac{C_5}{N_c} \right)
+\frac{3}{2}e_q\left( C_{8}+\frac{C_7}{N_c} \right) \;,
\nonumber\\
a_2^{T}(t)&=&
C_2+\frac{C_1}{N_c}\;,
\end{eqnarray}
where $a_2^{T}$ is the Wilson coefficient for the amplitude $F_a^T$.
The hard functions, which are the Fourier transformation of the virtual
quark propagator and the hard gluon propagator, are given by
\begin{eqnarray}
h_e(x_1,x_2,b_1,b_2)&=&K_{0}\left(\sqrt{x_1x_2}M_Bb_1\right)
\left[\theta(b_1-b_2)K_0\left(\sqrt{x_2}M_B
b_1\right)I_0\left(\sqrt{x_2}M_Bb_2\right)\right.
\nonumber \\
& & \hspace{33mm} \left.+\theta(b_2-b_1)K_0\left(\sqrt{x_2}M_Bb_2\right)
I_0\left(\sqrt{x_2}M_Bb_1\right)\right]\;,\\
h_a(x_2,x_3,b_2,b_3)&=&\left(\frac{i\pi}{2}\right)^2
H_0^{(1)}\left(\sqrt{x_2x_3}M_Bb_2\right)
\left[\theta(b_2-b_3)
H_0^{(1)}\left(\sqrt{x_3}M_Bb_2\right)
J_0\left(\sqrt{x_3}M_Bb_3\right)\right.
\nonumber \\
& & \hspace{41mm} \left.+\theta(b_3-b_2)H_0^{(1)}
\left(\sqrt{x_3}M_Bb_3\right)
J_0\left(\sqrt{x_3}M_Bb_2\right)\right]\;.
\end{eqnarray}

\subsection{Nonfactorizable Amplitudes}\label{sec:nfa}
%%%%%%%%%%%%%%% Fig. 5 %%%%%%%%%%%%%%%%%%%%%%%%%%%%%%%%%%%%%%%
\begin{figure}[hbt]
\begin{center}
\includegraphics{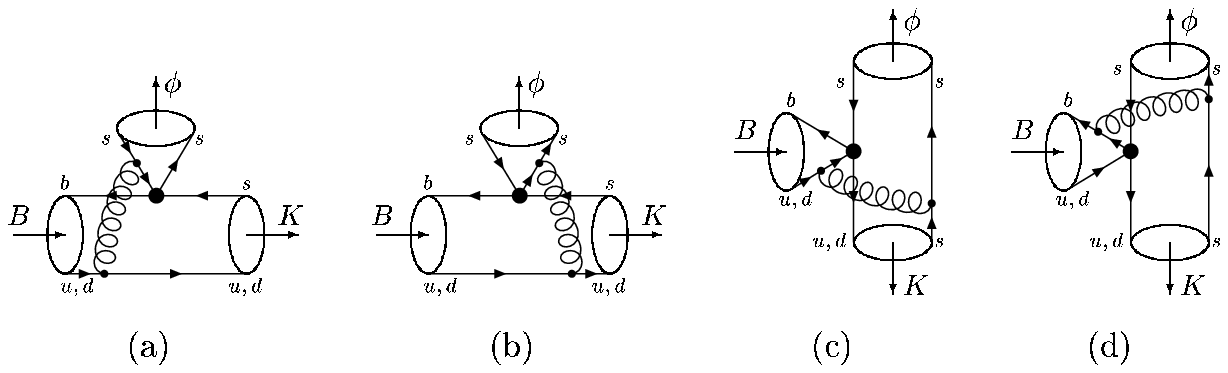}
\end{center}
\caption{Feynman diagrams contributing to nonfactorizable amplitudes}
\label{fig:nonfact}
\end{figure}
%%%%%%%%%%%%%%%%%%%%%%%%%%%%%%%%%%%%%%%%%%%%%%%%%%%%%%%%%%%%%%
The nonfactorizable amplitudes which come from
Fig.~\ref{fig:nonfact} are written as
\begin{eqnarray}
{\cal M}^P_e&=&{\cal M}^P_{e4}+{\cal M}^P_{e5}+{\cal M}^P_{e6}\;,
\\
{\cal M}^P_{e4}&=&
16\pi C_F \frac{\sqrt{2N_c}}{N_c}M_B^4\int_0^1 [dx]\int_0^{\infty}
b_1 db_1 b_3 db_3\phi_B(x_1,b_1)\phi_{\phi}(x_3)
\nonumber \\
& &\times 
\bigg\{
\left[(1-x_1-x_3)\phi_K^A(x_2)
-r_K x_2 \left( \phi^P_K(x_2) - \phi^T_K(x_2) \right) \right]
E'_{e4}(t^{(1)}_d)
h^{(1)}_d(x_1,x_2,x_3,b_1,b_1,b_3)
\nonumber \\
& &\;\;\;\;\;\; + \left[(x_1-x_2-x_3)\phi_K^A(x_2)
+r_K x_2 \left(  \phi_K^P(x_2)+\phi_K^T(x_2) \right) \right]
E'_{e4}(t^{(2)}_d)
h^{(2)}_d(x_1,x_2,x_3,b_1,b_1,b_3)
\bigg\}\;,
\\
{\cal M}^P_{e5}&=&
16\pi C_F \frac{\sqrt{2N_c}}{N_c}M_B^4\int_0^1 [dx]\int_0^{\infty}
b_1 db_1 b_3 db_3\phi_B(x_1,b_1)\phi_{\phi}(x_3)
\nonumber \\
& &\times 
\bigg\{
\left[(1-x_1+x_2-x_3)\phi_K^A(x_2)
-r_K x_2 \left( \phi^P_K(x_2) + \phi^T_K(x_2) \right) \right]
E'_{e5}(t^{(1)}_d)
h^{(1)}_d(x_1,x_2,x_3,b_1,b_1,b_3)
\nonumber\\
& & \;\;\;\;\;\;
+\left[(x_1-x_3)\phi_K^A(x_2)
+r_K x_2 \left(  \phi_K^P(x_2)-\phi_K^T(x_2) \right) \right]
E'_{e5}(t^{(2)}_d)
h^{(2)}_d(x_1,x_2,x_3,b_1,b_1,b_3)
\bigg\}\;,
\\
{\cal M}^P_{e6}&=&
16\pi C_F \frac{\sqrt{2N_c}}{N_c}M_B^4 r_{\phi} 
\int_0^1 [dx]\int_0^{\infty}
b_1 db_1 b_3 db_3\phi_B(x_1,b_1)
\nonumber \\
& &\times 
\bigg\{
\left[
(1-x_1-x_3)\phi_K^A(x_2) 
\left( \phi_{\phi}^t(x_3) + \phi_{\phi}^s(x_3) \right)
\right.\nonumber \\
& &\;\;\;\;\;\;\left.
+r_K(1-x_1-x_3)\left(  \phi_K^P(x_2)-\phi_K^T(x_2) \right) 
\left( \phi_{\phi}^t(x_3) + \phi_{\phi}^s(x_3) \right)
\right.\nonumber \\
& &\;\;\;\;\;\;\left.
-r_K x_2 \left(  \phi_K^P(x_2)+\phi_K^T(x_2) \right) 
\left( \phi_{\phi}^t(x_3) - \phi_{\phi}^s(x_3) \right)
\right]
E'_{e6}(t^{(1)}_d)
h^{(1)}_d(x_1,x_2,x_3,b_1,b_1,b_3)
\nonumber \\
& &
- \left[
(x_1-x_3)\phi_K^A(x_2) 
\left( \phi_{\phi}^t(x_3) - \phi_{\phi}^s(x_3) \right)
+r_K(x_1-x_3)
\left(  \phi_K^P(x_2)-\phi_K^T(x_2) \right) 
\left( \phi_{\phi}^t(x_3) - \phi_{\phi}^s(x_3) \right)
\right.\nonumber \\
& &\;\;\;\;\;\;\left.
+r_K x_2 \left(  \phi_K^P(x_2)+\phi_K^T(x_2) \right) 
\left( \phi_{\phi}^t(x_3) + \phi_{\phi}^s(x_3) \right)
\right]
E'_{e6}(t^{(2)}_d)
h^{(2)}_d(x_1,x_2,x_3,b_1,b_1,b_3)
\bigg\}\;.
\end{eqnarray}
${\cal M}^P_{e3}$ is added to ${\cal M}^P_{e4}$ term, because ${\cal
M}^P_{e3}$ is the same as ${\cal M}^P_{e4}$ without the Wilson
coefficient.  
\begin{eqnarray}
{\cal M}^P_a&=&{\cal M}^P_{a4}+{\cal M}^P_{a6}\;,
\\
{\cal M}^P_{a4}&=&
16\pi C_F \frac{\sqrt{2N_c}}{N_c}M_B^4\int_0^1 [dx]\int_0^{\infty}
b_1 db_1 b_3 db_3\phi_B(x_1,b_1)
\nonumber \\
& &\times 
\bigg\{
\left[
(1-x_2)\phi_K^A(x_2)\phi_{\phi}(x_3)
+r_Kr_\phi
\left(
(1-x_2)\left(  \phi_K^P(x_2)-\phi_K^T(x_2) \right) 
\left( \phi_{\phi}^t(x_3) + \phi_{\phi}^s(x_3) \right)
\right.\right.\nonumber \\
& &\;\;\;\;\;\;\left.\left.
+(x_1-x_3)\left(  \phi_K^P(x_2)+\phi_K^T(x_2) \right) 
\left( \phi_{\phi}^t(x_3) - \phi_{\phi}^s(x_3) \right)
\right)
\right]
E'_{a4}(t^{(1)}_f)
h^{(1)}_f(x_1,x_2,x_3,b_1,b_3,b_3)
\nonumber \\
& &
- \left[
(x_1+x_3)\phi_K^A(x_2)\phi_{\phi}(x_3)
\right.\nonumber \\
& &\;\;\;\;\;\;\left.
+r_Kr_\phi
\left(
4\phi_K^P(x_2)\phi_{\phi}^s(x_3)-(1-x_1-x_3)
\left(  \phi_K^P(x_2)-\phi_K^T(x_2) \right) 
\left( \phi_{\phi}^t(x_3) + \phi_{\phi}^s(x_3) \right)
\right.\right.\nonumber \\
& &\;\;\;\;\;\;\left.\left.
+x_2\left(  \phi_K^P(x_2)+\phi_K^T(x_2) \right) 
\left( \phi_{\phi}^t(x_3) - \phi_{\phi}^s(x_3) \right)
\right)
\right]
E'_{a4}(t^{(2)}_f)
h^{(2)}_f(x_1,x_2,x_3,b_1,b_3,b_3)
\bigg\}\;,\\
{\cal M}^P_{a6}&=&
16\pi C_F \frac{\sqrt{2N_c}}{N_c}M_B^4\int_0^1 [dx]\int_0^{\infty}
b_1 db_1 b_3 db_3\phi_B(x_1,b_1)
\nonumber \\
& &\times 
\bigg\{
\Big[ -r_K (1-x_2) \left( \phi^P_K(x_2) - \phi^T_K(x_2) \right)
\phi_{\phi}(x_3)
\nonumber\\
& & \;\;\;\;\;\;
-r_{\phi} (x_1-x_3) \phi^A_K(x_2) 
\left( \phi_{\phi}^t(x_3) + \phi_{\phi}^s(x_3) \right) 
\Big]
E'_{a6}(t^{(1)}_f)
h^{(1)}_f(x_1,x_2,x_3,b_1,b_3,b_3)
\nonumber \\
& & + \Big[ -r_K (1+x_2) \left( \phi^P_K(x_2) - \phi^T_K(x_2) \right)
\phi_{\phi}(x_3)
\nonumber\\
& & \;\;\;\;\;\;
-r_{\phi} (-2+x_1+x_3) \phi^A_K(x_2) 
\left( \phi_{\phi}^t(x_3) + \phi_{\phi}^s(x_3) \right) 
 \Big]
E'_{a6}(t^{(2)}_f)
h^{(2)}_f(x_1,x_2,x_3,b_1,b_3,b_3)
\bigg\}\;.
\end{eqnarray}
The tree amplitude ${\cal M}^T_a$ is the same as ${\cal M}_{a4}$, but
the Wilson coefficient is different. 
The evolution factors are defined by 
\begin{eqnarray}
E'_{ei}(t)&=&\alpha_s(t)a'_{ei}(t)\exp[-S(t)|_{b_2=b_1}]\;,
\;\;
E'_{ai}(t)=\alpha_s(t)a'_{ai}(t)\exp[-S(t)|_{b_2=b_3}]\;.
\end{eqnarray}
The hard scales $t$ are defined by 
\begin{eqnarray}
t^{(i)}_d&=&{\rm max}(DM_B,\sqrt{|D_i^2|}M_B,1/b_1,1/b_3)\;,
\;\;
t^{(i)}_f={\rm max}(FM_B,\sqrt{|F_i^2|}M_B,1/b_1,1/b_3)\;.
\end{eqnarray}
The Wilson coefficients are given by
\begin{eqnarray}
a'_{e4}(t) &=& C_3+C_4-\frac{1}{2}\left( C_9+C_{10} \right)\;,
\;\;\;\;\;
a'_{e5}(t) = C_6-\frac{1}{2}C_8\;,\nonumber\\ 
a'_{e6}(t) &=& C_5-\frac{1}{2}C_7\;,
\;\;\;\;\;\;\;\;\;\;\;\;\;\;\;\;\;\;\;\;\;\;\;\;\;\;\;
a'_{a4}(t) = C_3+\frac{3}{2}e_qC_9\;,\nonumber\\
a'_{a6}(t) &=& C_5+\frac{3}{2}e_qC_7\;,
\;\;\;\;\;\;\;\;\;\;\;\;\;\;\;\;\;\;\;\;\;\;\;\;\;
a^T(t) = C_1.
\end{eqnarray}
The hard functions are given by
\begin{eqnarray}
h^{(j)}_d&=& \left[\theta(b_1-b_3)K_0\left(DM_B
b_1\right)I_0\left(DM_Bb_3\right)
+\theta(b_3-b_1)K_0\left(DM_B b_3\right)
I_0\left(DM_B b_1\right)\right]  
\nonumber \\
&  & \hspace{10mm} \times  K_{0}(D_{j}M_Bb_{3})
\;\;\;\;\;\;\;\;\;\;\;\;\;\;\;\;\;  
\mbox{for $D^2_{j} \geq 0$}
\nonumber  \\
&  & \hspace{10mm} \times \frac{i\pi}{2} H_{0}^{(1)}
(\sqrt{|D_{j}^2|}M_Bb_{3})\;\;\;\;
\mbox{for $D^2_{j} \leq 0$}\;,
\\
h^{(j)}_f&=& \frac{i\pi}{2}
\left[\theta(b_1-b_3)H_0^{(1)}\left(FM_B
b_1\right)J_0\left(FM_Bb_3\right)
+\theta(b_3-b_1)H_0^{(1)}\left(FM_B b_3\right)
J_0\left(FM_B b_1\right)\right]
\nonumber \\
&  & \hspace{10mm} \times
 K_{0}(F_{j}M_Bb_{1})\;\;\;\;\;\;\;\;\;\;\;\;\;\;\;\;\;  
\mbox{for $F^2_{j} \geq 0$} 
\nonumber\\
&  & \hspace{10mm} \times \frac{i\pi}{2} H_{0}^{(1)}
(\sqrt{|F_{j}^2|}M_Bb_{1})
\;\;\;\; \mbox{for $F^2_{j} \leq 0$}\;,
\end{eqnarray}
with the variables
\begin{eqnarray}
D^{2}&=&x_{1}x_{2}\;,\;\;\;\;\;\;\;\;\;\;\;\;\;\;\;\;\;\;\;\;\;\;\;\;
F^{2}=(1-x_{2})x_{3}\;,\nonumber \\
D_{1}^{2}&=&\{x_{1}-(1-x_{3})\}x_{2}\;,\;\;\;\;
F_{1}^{2}=(x_1-x_3)(1-x_2)\;,\nonumber \\
D_{2}^{2}&=&(x_{1}-x_{3})x_{2}\;,\;\;\;\;\;\;\;\;\;\;\;\;\;\;
F_{2}^{2}=x_{1}+x_{3}+(1-x_{1}-x_{3})(1-x_{2})\;.
\end{eqnarray}

%%%%%%%%%%%%%%%%%%%%%%%%%%%%%%%%%%%%%%%%%%%%%%%%%%%%%%
%
%            Numerical Results
%
%%%%%%%%%%%%%%%%%%%%%%%%%%%%%%%%%%%%%%%%%%%%%%%%%%%%%%

\section{Numerical Results}\label{sec:num}
We summarize the parameters which is used in the numerical
analysis of our calculation.

We use the model of the $B$ meson wave function written as
\begin{eqnarray}
\phi_B(x,b) &=& N_B
x^2 (1-x)^2\exp\left[-\frac{1}{2}
\left(\frac{xM_B}{\omega_{B}}\right)^2
-\frac{\omega_{B}^2 b^2}{2}\right] \;,
\end{eqnarray}
where $\omega_{B}=0.40$ GeV\cite{BW}. $N_B$ is determined
from normalization condition Eq.~(\ref{eq:bnor}). 

The $K$ meson wave functions are given as
\begin{eqnarray}
\phi_{K}^A(x) &=& \frac{f_K}{2\sqrt{2N_c}}6x(1-x)
\left[1 + 3a_1(1-2x)
+ \frac{3}{2}a_2\{5(1-2x)^2-1\}\right] \;,
\\
\phi^P_{K}(x) &=& \frac{f_K}{2\sqrt{2N_c}}
\bigg[ 
1
+\frac{1}{2}\left(30\eta_3 -\frac{5}{2}\, \rho_K^2\right)
\left\{3(1-2x)^2-1\right\} \nonumber\\
& & -\frac{1}{8}\left( 3\eta_3\omega_3+\frac{27}{20}\, \rho_K^2
  + \frac{81}{10}\, \rho_K^2 a_2\right)
\left\{3-30(1-2x)^2+35(1-2x)^4\right\} 
\bigg]\;,
\\
\phi^T_{K}(x) &=& \frac{f_K}{2\sqrt{2N_c}}(1-2x)
\bigg[ 1
+
6\left(5\eta_3 -\frac{1}{2}\,\eta_3\omega_3 - \frac{7}{20}\,
      \rho_K^2 - \frac{3}{5}\,\rho_K^2 a_2 \right)
(1-10x+10x^2) \bigg]\;.
\end{eqnarray}
where $\rho_K=(M_d+m_s)/M_K$\cite{PB1}\cite{PB2}.
The parameters of these wave functions are given as 
$a_1=0.17,\;
a_2=0.20, \;
\eta_3=0.015,\; 
\omega_3=-3.0$ where renormalization scale is 1 GeV. We fix the
parameters as above values. 

The $\phi$ meson wave functions are given as
\begin{eqnarray}
\phi_{\phi}(x) &=& \frac{f_{\phi}}{2\sqrt{2N_c}}6x(1-x)\;,
\\
\phi^t_{\phi}(x) &=& \frac{f^T_{\phi}}{2\sqrt{2N_c}}
\bigg[ 3(1-2x)^2
+\frac{35}{4}\zeta_3^T\{3-30(1-2x)^2+35(1-2x)^4\} \nonumber\\
& & +\frac{3}{2}\delta_{+}\left\{1-(1-2x)\log\frac{1-x}{x}\right\} 
\bigg]\;,
\\
\phi^s_{\phi}(x) &=& \frac{f^T_{\phi}}{4\sqrt{2N_c}}
\bigg[ 
(1-2x)\left(
6+9\delta_{+}+140\zeta_3^T-1400\zeta_3^Tx+1400\zeta_3^Tx^2
\right)+3\delta_{+}\log\frac{x}{1-x}
 \bigg]\;,
\end{eqnarray}
where $\zeta_3^T=0.024,\; \delta_{+}=0.46$~\cite{PB3}. 
We fix the parameters as above values, since the numerical results 
is insensitive to these parameters. 

We use the Wolfenstein parameters for the CKM matrix elements   
$A = 0.819, \;
\lambda = 0.2196, \;
R_b \equiv \sqrt{\rho^2+\eta^2}= 0.38$~\cite{PDG}, and choose the angle
$\phi_3 = \pi/2$~\cite{KLS}. 
In addition, we use the following parameters.
\begin{center}
$M_B = 5.28\; {\rm GeV},\;\;\;
M_K = 0.49 \;{\rm GeV},\;\;\;
M_{\phi}=1.02\; {\rm GeV},\;\;\; 
m_{0K} =1.70\; {\rm GeV}$,\\
$f_{B} = 190\; {\rm MeV},\;\;\;
f_{K} = 160\; {\rm MeV},\;\;\;
f_{\phi} = 237 \;{\rm MeV},\;\;\;
f_{\phi}^T = 215 \;{\rm MeV}$,\\
$\tau_{B^0}=1.55\times 10^{-12}\;{\rm sec},\;\;\;
 \tau_{B^\pm}=1.65\times 10^{-12} \;{\rm sec} \;,\;\;\;$
$\Lambda_{\rm QCD}^{(4)}=0.250\; {\rm GeV}$.\\
\end{center}

We show the numerical results of each amplitude for the $B^0 \to \phi
K^0$ and $B^\pm \to \phi K^\pm$ decays in Table \ref{table1}.
The factorizable penguin amplitude $F_e^P$ is
dominant contribution to the $B \to \phi K$ decays.
The factorizable annihilation penguin amplitude $F_a^P$ generates large
strong phase.
In the $B^\pm \to \phi K^\pm$ modes, there are the contribution from
$f_B F_{a}^T$ and $M_{a}^T$. These tree amplitudes contribute a few percent
to the whole amplitude, since the CKM matrix elements related to the
tree amplitudes are very small. 

We predict the branching ratios for the $B \to \phi K$ decays as
\begin{eqnarray}
{\rm Br}(B^0\to \phi K^0)&=& 
\left| \frac{f_B f_K}{190\; {\rm MeV}\; 160\; {\rm MeV}} \right|^2
\times \left( 9.43\times 10^{-6} \right)\;,   \\
{\rm Br}(B^\pm\to \phi K^\pm)&=& 
\left| \frac{f_B f_K}{190\; {\rm MeV}\; 160\; {\rm MeV}} \right|^2
\times \left( 10.1\times 10^{-6} \right)\;.
\end{eqnarray}
The current experimental values are summarized in Table \ref{table2}. 
The values which are predicted in PQCD are consistent with the current
experimental data. 
However, our branching ratios have the theoretical error from the
$O(\alpha_s^2)$ corrections, the higher twist corrections, and the error
of input parameters. 
Large uncertainties come from the meson decay constants, the
shape parameter 
$\omega_B$, and $m_{0K}$. These parameters are fixed from the other
modes($B\to K\pi,\; D\pi \; \pi\pi$, etc.).  
We try to vary $\omega_B$ from 0.36 GeV to 0.44 GeV, then we obtain 
Br($B^\pm\to \phi K^\pm) = (7.54 \sim 13.9)\times 10^{-6}$. 
Next, we set $\omega_B=0.40$ and try to vary
$m_{0K}$ from 1.40 GeV to 1.80 GeV, then we obtain  Br($B^\pm
\to \phi K^\pm) = (6.65 \sim 11.4)\times 10^{-6}$.

In the naive factorization assumption, the branching ratio is very 
sensitive to the effective number of colors $N_c^{\rm eff}$. If we set
$N_c^{\rm eff}$=3, then the branching ratio is about $4.5\times 10^{-6}$
where the scale of the Wilson coefficient is taken to $M_B/2$ and
$F^{BK}$ is 0.38 from the BSW model. 
Our predicted values are larger than those obtains from the naive
factorization assumption. 
This is due to the enhancement of the Wilson coefficient. In PQCD
approach, the scale  of the Wilson coefficients, which is equal to the
hard scale $t$, can reach lower values than $M_B/2$ in 
case of the naive factorization assumption,
and the Wilson coefficients are then enhanced\cite{KLS}. Therefore, our
branching ratios are enhanced over those
obtained from the naive factorization assumption. 

In the QCD-improved factorization, the branching ratios for the $B \to
\phi K$ decays are predicted as Br$(B^0 \to \phi K^0) =
(4.0^{+2.9}_{-1.4})\times 10^{-6}$ and Br$(B^- \to \phi K^-) =
(4.3^{+3.0}_{-1.4})\times 10^{-6}$ with annihilation
effect\cite{QIF2}. Our predicted values are larger than these values. 

We also compute the direct CP asymmetry defined as
\begin{eqnarray}
A_{CP}^{\rm dir}=\frac{|A^+|^2-|A^-|^2}{|A^+|^2+|A^-|^2}\label{eq:acp}\;.
\end{eqnarray}
If the amplitude of the $B^\pm \to \phi K^\pm$ decay can be written as 
\begin{eqnarray}
A^+ &=& V_{ts}V_{tb}^*e^{i\delta_P}A_P-V_{us}V_{ub}^*e^{i\delta_T}A_T\;,\\
A^- &=& V_{ts}^*V_{tb}e^{i\delta_P}A_P-V_{us}^*V_{ub}e^{i\delta_T}A_T\;,
\end{eqnarray}
where the indices $T$ and $P$ denote the tree and penguin respectively,
$\delta_T$ and $\delta_P$ are the strong phase, and the amplitude $A_T$
and $A_P$ are real, then the direct CP asymmetry is given by 
\begin{eqnarray}
A_{CP}^{\rm dir}=\frac{2\lambda^2R_b\sin\phi_3\sin(\delta_P-\delta_T)
\frac{A_T}{A_P}}
{1+\lambda^4R_b \left(\frac{A_T}{A_P}\right)^2}\;.
\end{eqnarray}
Our results of the direct CP asymmetry is about few percent.
The reason why the direct CP asymmetry is very small is that the CKM
matrix element for the tree amplitude is much smaller than for the
penguin amplitude.

%%%%%%%%%%%%%%%%%%%%%%%%%%%%%%%%%%%%%%%%%%%%%%%%%%%%%%
%
%            Summary
%
%%%%%%%%%%%%%%%%%%%%%%%%%%%%%%%%%%%%%%%%%%%%%%%%%%%%%%

\section{Summary}
In this paper, we calculate the $B^0\to \phi K^0$ and $B^\pm\to \phi
K^\pm$ decays in PQCD approach. Our predicted branching ratios agree
with the current experimental data and are larger than the values
obtained by the naive factorization assumption and the QCD-improved
factorization. This is because the Wilson coefficients related to
penguin operators are enhanced dynamically in PQCD.

In the PQCD approach, 
we can calculate the strong phase from nonfactorizable amplitudes and
annihilation amplitudes explicitly. We can then calculate the direct CP
asymmetry of this modes, however this asymmetry is very small since the
CKM matrix element for the tree amplitude is much smaller than for the
penguin amplitude.

%%%%%%%%%%%%%%%%%%%%%%%%%%%%%%%%%%%%%%%%%%%%%%%%%%%%%%%%%%%%%%%%

\vspace{0.5cm}
\noindent{\bf Note added:}
           
After this work has been completed, we become aware of a similar
calculation by Chen, Keum and Li\cite{CKL}.
Our results are in agreement.

%%%%%%%%%%%%%%%%%%%%%%%%%%%%%%%%%%%%%%%%%%%%%%%%%%%%%%%%%%%%%%%%

\vspace{0.5cm}
\centerline{\bf Acknowledgment}
\vspace{0.5cm}

The topic of this research was suggested by Professor A.I. Sanda. 
The author thanks our PQCD group members: Y.Y. Keum, E. Kou, T. Kurimoto,
H-n. Li, T. Morozumi, R. Sinha, K. Ukai for useful discussions.
The author thanks JSPS for partial support.
The work was supported in part by Grant-in Aid for Special Project
Research (Physics of CP violation); Grant-in Aid for Scientific Research
from the Ministry of Education, Science and Culture of Japan.

%%%%%%%%%%%%%%%%%%%%%%%%%%%%%%%%%%%%%%%%%%%%%%%%%%%%%%%%%%%%%%%%%
%
%              Appendix
%
%%%%%%%%%%%%%%%%%%%%%%%%%%%%%%%%%%%%%%%%%%%%%%%%%%%%%%%%%%%%%%%%%
\appendix

\section{Wilson coefficients}
The Wilson coefficients for the $m_b=4.8$ GeV are 
\begin{eqnarray}
C_1(m_b) &=& -2.693 \times 10^{-1}\;, \hspace{7mm}
C_2(m_b) = 1.123\;, \nonumber \\
C_3(m_b) &=& 1.243 \times 10^{-2}\;, \hspace{10mm} 
C_4(m_b) = -2.665 \times 10^{-2}\;, \hspace{10mm}  \nonumber \\
C_5(m_b) &=& 7.753 \times 10^{-3}\;, \hspace{10mm} 
C_6(m_b) = -3.256 \times 10^{-2}\;, \nonumber \\
C_7(m_b) &=& 3.426 \times 10^{-4}\;, \hspace{10mm} 
C_8(m_b) = 3.131 \times 10^{-4}\;, \hspace{10mm} \nonumber \\
C_9(m_b) &=& -9.732 \times 10^{-3}\;, \hspace{6mm} 
C_{10}(m_b) = 2.217 \times 10^{-3}\;.
\end{eqnarray}

\section{$B$ Meson Wave Functions}
We consider the $B$ meson wave function
$\Phi^{(in)}_{B,\alpha\beta,ij}=\mat{0}{{\bar b}_{\beta j}(0)d_{\alpha
i}(z)}{B(P_1)}$ where the indices $\alpha$ and $\beta$ are the spin
indices, $i$ and $j$ are the color indices. This wave function can be
decomposed into some components with different spin structure.  
The $B$ meson wave function is then decomposed to be 
\begin{eqnarray}
\Phi^{(in)}_{B,\alpha\beta,ij}
&=&
\frac{i \delta_{ij}}{\sqrt{2N_c}}\int dx_1 d^2{{\bf k}_{1T}}
 e^{-i(x_1P_1^-z^+-{{\bf k}_{1T}}{{\bf z}_T})}
\left[
(\not P_1 \gamma_5)\phi_B^A(x_1,{\bf k}_{1T})
+\gamma_5 \phi_B^P(x_1,{\bf k}_{1T})
\right]_{\alpha\beta}\;,
\end{eqnarray}
where $N_c$ is color's degree of freedom.
The axial-vector wave function $\phi_B^A(x_1,{\bf k}_{1T})$ is defined by 
\begin{eqnarray}
P_{1\mu}\int dx_1 d^2{{\bf k}_{1T}} 
e^{-i(x_1P_1^-z^+-{{\bf k}_{1T}}{{\bf z}_T})}
\phi_B^A(x_1,{\bf k}_{1T})
=\frac{ -i }{2\sqrt{2N_c}}
\mat{0}{{\bar b}(0) \gamma_5\gamma_\mu d(z)}{\bar b(P_1)}\;,
\end{eqnarray}
and the pseudo-scalar wave function $\phi_B^P(x_1,{\bf k}_{1T})$ is
defined by  
\begin{eqnarray}
\int dx_1 d^2{{\bf k}_{1T}} e^{-i(x_1P_1^-z^+-{{\bf k}_{1T}}{{\bf z}_T})}
\phi_B^P(x_1,{\bf k}_{1T})
=\frac{ -i }{2\sqrt{2N_c}}\mat{0}{{\bar b}(0) \gamma_5 b(z)}{\bar d(P_1)}\;.
\end{eqnarray}
Using HQET, we can denote $\phi_B^P(x_1,{\bf
k}_{1T})=M_B\phi_B^A(x_1,{\bf k}_{1T})$. The $B$ meson wave function is 
then written by 
\begin{eqnarray}
\Phi^{(in)}_{B,\alpha\beta,ij}&=&
\frac{i \delta_{ij}}{\sqrt{2N_c}}\int dx_1 d^2{{\bf k}_{1T}} 
e^{-i(x_1P_1^-z^+-{{\bf k}_{1T}}{{\bf z}_T})}
\left[
(\not P_1+M_B)\gamma_5 \phi_B(x_1,{\bf k}_{1T})
\right]_{\alpha\beta}\;.
\end{eqnarray}
The normalization of this wave function is given by
\begin{eqnarray}
\int_0^1 dx_{1}\phi_B(x_{1},b_1=0)&=&\frac{f_B}{2\sqrt{2N_c}}\;,
\label{eq:bnor}
\end{eqnarray}
where $b_1$ is the conjugate space of $k_1$, and $f_B$ is the decay
constant of the $B$ meson.

\section{$K$ Meson Wave Functions}
The $K$ meson wave function is defined that
$\Phi^{(in)}_{K,\alpha\beta,ij}=\mat{0}{{\bar s}_{\beta j}(0)d_{\alpha
i}(z)}{K(P_2)}$, $z=(0,z^-,{\bf 0}_T)$ is the coordinate of the $d$ quark.
$\Phi^{(in)}_{K,\alpha\beta,ij}$ can be decomposed into some components with
different spin structure $\gamma_5\gamma_\mu$, $\gamma_5$,
$\gamma_5\sigma_{\mu\nu}$, $\cdots$. These matrix elements are
calculated using the twist expansion in the light-cone QCD sum
rules. The twist expansion is the expansion in terms of $m_{0K}$ where
$m_{0K}= M_K^2/(m_d+m_s)$. The leading twist terms are independent on
$m_{0K}$, the twist-3 and twist-4 terms are proportional to $m_{0K}$ and
$m_{0K}^2$ respectively. Up to twist-3, the matrix elements are given
as\cite{PB1}\cite{PB2}  
\begin{eqnarray}
\langle 0|{\bar s}(0)\gamma_5\gamma_\mu d(z)|K(P_2)\rangle&=&
-if_K P_{2\mu}\int_0^1 dx_2 e^{-ix_2P_2\cdot z}\phi_v(x_2)\;,
\\
\langle 0|{\bar s}(0)\gamma_5 d(z)|K(P_2)\rangle&=&
-if_K m_{0K}\int_0^1 dx_2 e^{-ix_2P_2\cdot z}\phi_p(x_2)\;,
\\
\langle 0|{\bar s}(0)\gamma_5\sigma_{\mu\nu}d(z)|K(P_2)\rangle
&=&
\frac{i}{6} f_Km_{0K}\left( 1-\frac{M_K^2}{m_{0K}^2} \right)
(P_{2\mu}z_\nu-P_{2\nu}z_\mu)
\int_0^1 dx_2 e^{-ix_2P_2\cdot z}\phi_\sigma(x_2)
\nonumber\\
&=&
\frac{1}{6} f_Km_{0K}
(v_{\mu}n_\nu-v_{\nu}n_\mu)
\int_0^1 dx_2 e^{-ix_2P_2\cdot z}\frac{d}{dx_2}\phi_\sigma(x_2)\;,
\end{eqnarray}
where  $v^\mu = P_2^\mu/P_2^+$ and
$n^\mu = z^\mu/z^-=(0,1,{\bf 0}_T)$.
$\phi_v(x_2)$ is the leading twist wave function, $\phi_p(x_2)$ and
$d/dx_2\phi_\sigma(x_2)$ are the twist-3 wave functions. 
These wave functions are normalized as
\begin{eqnarray}
\int_0^1 dx_2\phi_v(x_2) = 1\;,\;
\int_0^1 dx_2\phi_p(x_2) = 1\;,\;
\int_0^1 dx_2 \frac{d}{dx_2}\phi_{\sigma}(x_2) = 0\;.
\end{eqnarray}
We define the following new wave functions,
\begin{eqnarray}
\phi_K^A(x_2)=\frac{f_K}{2\sqrt{2N_c}}\phi_{v}(x_2)\;,\;
\phi_K^P(x_2)=\frac{f_K}{2\sqrt{2N_c}}\phi_p(x_2)\;,\;
\phi_K^T(x_2)=\frac{f_K}{12\sqrt{2N_c}}
\frac{d}{dx_2}\phi_\sigma(x_2)\;.
\end{eqnarray}
The $K$ meson wave functions are then written as
\begin{eqnarray}
\Phi^{(in)}_{K,\alpha\beta,ij}
&=&\frac{-i\delta_{ij}}{\sqrt{2N_c}}\int^1_0dx_2 e^{-ix_2P_2\cdot z}
\left[
\not P_2\gamma_5\phi^A_K(x_2)+m_{0K}\gamma_5\phi_K^P(x_2)
+m_{0K} (\not n\not v -1)\gamma_5\phi_K^T(x_2)
\right]_{\alpha\beta}\;.
\end{eqnarray}

\section{$\phi$ Meson Wave Functions}
The $\phi$ meson wave function has much terms than the $B$ meson and $K$
meson wave functions, since the $\phi$ meson is a vector meson. 
The wave function $\Phi^{(in)}_{\phi,\alpha\beta,ij}=\mat{0}{{\bar
s}_{\beta j}(0)s_{\alpha i}(z)}{\phi(P_3)}$ can be decomposed into some
components with different spin structure $\gamma_\mu$,
$\gamma_\mu\gamma_5$, $\sigma_{\mu\nu}$, $I$, $\cdots$. The components
are calculated using the twist expansion in the light-cone QCD sum
rules. In this case, the twist expansion is the expansion in terms of
$M_\phi$. Up to twist-3, the components are given as\cite{PB3} 
\begin{eqnarray}
\langle 0|{\bar s}(0)\gamma_\mu s(z)|\phi(P_3)\rangle
&=&
f_{\phi} M_{\phi}
\int_0^1 dx_3 e^{-ix_3P_3\cdot z}
\left[
P_{3\mu}\frac{\epsilon_\phi\cdot z}{P_3\cdot z}\phi_{//}(x_3)
+\epsilon^T_{\phi\mu}g^{(v)}_{\perp}(x_3)
\right]
\nonumber\\
&=&
f_{\phi} M_{\phi}
\int_0^1 dx_3 e^{-ix_3P_3\cdot z}
\left[
\epsilon_{\phi\mu}\phi_{//}(x_3)
+\epsilon^T_{\phi\mu}g^{(v)}_{\perp}(x_3)
\right]\;,
\\
\langle 0|{\bar s}(0)\gamma_\mu\gamma_5 s(z)|\phi(P_3)\rangle
&=&
-\frac{1}{4}\left(f_\phi-f^T_\phi\frac{2m_s}{M_\phi}\right)
M_\phi\epsilon_\mu^{\;\;\nu\alpha\beta}\epsilon^T_{\phi\nu}
P_{3\alpha}z_\beta
\int_0^1 dx_3 e^{-ix_3P_3\cdot z}g^{(a)}_{\perp}(x_3)
\nonumber\\
&=&
-\frac{1}{4}f_\phi
M_\phi\epsilon_\mu^{\;\;\nu\alpha\beta}\epsilon^T_{\phi\nu}
P_{3\alpha}z_\beta
\int_0^1 dx_3 e^{-ix_3P_3\cdot z}g^{(a)}_{\perp}(x_3)\;,
\\
\langle 0|{\bar s}(0)\sigma_{\mu\nu}s(z)|\phi(P_3)\rangle
&=&
i f^T_{\phi}\int_0^1 dx_3 e^{-ix_3P_3\cdot z}\bigg[
\left( 
\epsilon^T_{\phi\mu}P_{3\nu}-\epsilon^T_{\phi\nu}P_{3\mu} 
\right)
\phi_{\perp}(x_3)\nonumber\\
& &\;\;\;\;\;\;\;\;\;\;\;\;\;\;\;\;\;\;\;\;\;\;\;\;\;\;\;\;\;\;
+
(P_{3\mu} z_\nu-P_{3\nu} z_\mu)\frac{\epsilon_\phi\cdot z}
{(P_3\cdot z)^2}M_\phi^2
h^{(t)}_{//}(x_3)
\bigg]\nonumber\\
&=&
i f^T_{\phi}\int_0^1 dx_3 e^{-ix_3P_3\cdot z}\bigg[
\left( \epsilon^T_{\phi\mu}P_{3\nu}-\epsilon^T_{\phi\nu}P_{3\mu} \right)
\phi_{\perp}(x_3)\nonumber\\
& &\;\;\;\;\;\;\;\;\;\;\;\;\;\;\;\;\;\;\;\;\;\;\;\;\;\;\;\;\;\;
+
(\epsilon_{\phi\mu} P_{3\nu}-\epsilon_{\phi\nu} P_{3\mu})
h^{(t)}_{//}(x_3)
\bigg]\;,\\
\langle 0|{\bar s}(0)I s(z)|\phi(P_3)\rangle
&=&
\frac{i}{2} \left(f^T_\phi-f_\phi\frac{2m_s}{M_\phi}\right)
\epsilon_\phi\cdot z M^2_{\phi}
\int_0^1 dx_3 e^{-ix_3P_3\cdot z}
h^{(s)}_{//}(x_3)
\nonumber\\
&=&
- f_\phi^TM_{\phi}
\int_0^1 dx_3 e^{-ix_3P_3\cdot z}
\frac{1}{2}\frac{d}{dx_3}h^{(s)}_{//}(x_3)\;,
\end{eqnarray}
where $n_\mu= z_\mu/z^+$,$v_\mu=P_{3\mu}/P^-_3$.
$\phi_{//}(x_3)$ and $\phi_{\perp}(x_3)$ are the leading twist wave
functions, $h^{(t)}_{//}(x_3)$, $\frac{d}{dx_3}h^{(s)}_{//}(x_3)$,
$g^{(v)}_{\perp}(x_3)$ and $g^{(a)}_{\perp}(x_3)$ are the twist-3 wave
functions. 
These wave functions 
satisfy the normalization conditions
\begin{eqnarray}
\int_0^1 dx_3 \phi_{//}(x_3) = 1 \;,\;
\int_0^1 dx_3 h^{(t)}_{//}(x_3) =1 \;,\;
\int_0^1 dx_3 \frac{d}{dx_3}h^{(s)}_{//}(x_3)=0\;,\nonumber\\
\int_0^1 dx_3 \phi_{\perp}(x_3)= 1 \;,\;\;
\int_0^1 dx_3 g^{(v)}_{\perp}(x_3)= 1 \;,\;\;\;\;\;\;\;\;
\int_0^1 dx_3 g^{(a)}_{\perp}(x_3)= 1 \;.
\end{eqnarray}
We can neglect the wave functions which are proportional to the
transverse polarization vector $\epsilon_\phi^T$, because these terms do
not appear in our calculations. 
We define the following new wave functions,
\begin{eqnarray}
\phi_{\phi}(x_3)=\frac{f_{\phi}}{2\sqrt{2N_c}}\phi_{//}(x_3)\;,\;
\phi_{\phi}^t(x_3)=\frac{f^T_{\phi}}{2\sqrt{2N_c}}h^{(t)}_{//}(x_3)\;,\;
\phi_{\phi}^s(x_3)=\frac{f^T_{\phi}}{4\sqrt{2N_c}}
\frac{d}{dx_3}h^{(s)}_{//}(x_3)\;.
\end{eqnarray}
The $\phi$ meson wave functions are then written as
\begin{eqnarray}
\Phi^{(in)}_{\phi,\alpha\beta,ij}
&=&\frac{\delta_{ij}}{\sqrt{2N_c}}\int^1_0dx_3 e^{-ix_3P_3\cdot z}
\left[
M_\phi\not\epsilon_\phi \phi_\phi(x_3)+
\not P_3\not\epsilon_\phi\phi_\phi^t(x_3)
+ M_\phi\phi_\phi^s(x_3)
\right]_{\alpha\beta}\;.
\end{eqnarray}
It must be noted that this expression is dependent on how to define the
sign of $\epsilon_\phi$.

We considered the wave functions for an initial state. In this study, we
need the $K$ meson and $\phi$ meson wave functions for a final state. 
The wave function for a final state is given as
$\Phi^{(out)}=\gamma_0\Phi^{(in)\dagger}\gamma_0$.

%%%%%%%%%%%%%%%%%%%%%%%%%%%%%%%%%%%%%%%%%%%%%%%%%%%%%%%%%%%%%%%%%

\newpage
%%%%%%%%%%%% Table 1  %%%%%%%%%%%%%%%%%%%%%%%%%%%%%%%%%%%%%%%%%%%%
\begin{table}[hbt]
%\vspace*{0.5cm}
%\hspace*{-1.2cm}
\begin{center}
\begin{tabular}{ccc} 
&  $B^0 \to \phi K^0$ & $B^\pm \to \phi K^\pm$ 
\\
\hline
$f_\phi F_e^P$ & 
 $-1.03 \times 10^{-1}$ & 
 $-1.03 \times 10^{-1}$ 
\\
$f_B F_{a}^P$ & 
 $ 6.45 \times 10^{-3} + \,\, i \,\, 4.28 \times 10^{-2}$ &
 $ 6.17 \times 10^{-3} + \,\, i \,\, 4.20 \times 10^{-2}$ 
\\
$M_{e}^P$ & 
 $ 5.24 \times 10^{-3} - \,\, i \,\, 3.61 \times 10^{-3}$ &
 $ 5.24 \times 10^{-3} - \,\, i \,\, 3.61 \times 10^{-3}$ 
\\
$M_{a}^P$ & 
 $-8.03 \times 10^{-4} - \,\, i \,\, 1.73 \times 10^{-3}$ &      
 $-6.56 \times 10^{-4} - \,\, i \,\, 7.22 \times 10^{-4}$ 
\\
$f_B F_{a}^T$ & 
 &
 $-1.11 \times 10^{-1} - \,\, i \,\, 3.75 \times 10^{-2}$ 
\\
$M_{a}^T$ & 
 &
 $ 1.60 \times 10^{-2} + \,\, i \,\, 2.77 \times 10^{-2}$ 
\\
\end{tabular}
\end{center}
\caption{Contribution to the $B^0 \to \phi K^0$ and 
$B^\pm \to \phi K^\pm$ decays from each amplitude.}   
\label{table1}
\end{table}
%%%%%%%%%%%%%%%%%%%%%%%%%%%%%%%%%%%%%%%%%%%%%%%%%%%%%%%%%%%%%%%%%%
%%%%%%%%%%%% Table 2  %%%%%%%%%%%%%%%%%%%%%%%%%%%%%%%%%%%%%%%%%%%%
\begin{table}[hbt]
%\vspace*{0.5cm}
%\hspace*{-1.2cm}
\begin{center}
\begin{tabular}{ccc}
& ${\rm Br}(B^0\to \phi K^0)$ & ${\rm Br}(B^\pm\to \phi K^\pm)$
\\
\hline
BaBar &
$(8.1^{+3.1}_{-2.5}\pm 0.8)\times 10^{-6}$ &
$(7.7^{+1.6}_{-1.4}\pm 0.8)\times 10^{-6}$
\\
BELLE &
$(8.7^{+3.8}_{-3.0}\pm1.5)\times 10^{-6}$ &
$(10.6^{+2.1}_{-1.9}\pm2.2)\times10^{-6}$
\\
CLEO &
$ < 12.3 \times 10^{-6}$ &
$(5.5^{+2.1}_{-1.8}\pm 0.6)\times 10^{-6}$
\\
\end{tabular}
\end{center}
\caption{The experimental data of the $B \to \phi K$ branching ratios
 from the BaBar\protect\cite{BABAR}, the BELLE\protect\cite{BELLE} 
and the CLEO\protect\cite{CLEO}}    
\label{table2}
\end{table}
%%%%%%%%%%%%%%%%%%%%%%%%%%%%%%%%%%%%%%%%%%%%%%%%%%%%%%%%%%%%%%%%%%

%%%%%%%%%%%%%%%%%%%%%%%%%%%%%%%%%%%%%%%%%%%%%%%%%%%%%%%%%%%%%%%%%%%%
\end{document}